\documentclass[11pt]{article}
\usepackage{amsmath}
\usepackage{amsfonts}
\usepackage{amssymb}
\usepackage{graphicx}
\usepackage{fancyhdr}

\usepackage{color}

\begin{document}

\title{Holographic Q-Picture of Black Holes in Five Dimensional Minimal Supergravity}

\author{Chiang-Mei Chen$^1${\footnote{Email: cmchen@phy.ncu.edu.tw}},
V. Kamali$^2${\footnote{Email: vkamali1362@gmail.com}}, and
M. R. Setare$^2${\footnote{Email: rezakord@ipm.ir}}
\\
{\small $^1$Department of Physics and Center for Mathematics and Theoretical Physics,} \\
{\small National Central University, Chungli 320, Taiwan}
\\
{\small $^2$ Department of Science of Bijar, Kurdistan University, Bijar, Iran}}

\date{}
\maketitle

\begin{abstract}
In this article, we explore the holographic Q-picture description for the charged rotating black holes in the five-dimensional minimal supergravity. The central charge in the Q-picture depends only on black hole charge, therefore can be computed from the near horizon geometry of the extremal and non-rotating counterpart. Moreover, the CFT temperatures can be identified by studying the hidden conformal symmetry, and the related gravity-CFT dictionary can be translated via thermodynamics analysis. The entropy and absorption cross section computed from both gravity and CFT sides properly agree with each other.
\end{abstract}

\newpage

\section{Introduction}
The holographic principle~\cite{'tHooft:1993gx, Susskind:1994vu, Maldacena:1997re} is an outstanding concept which provides dual descriptions connecting gravity and field theory. In the past years, numerous substantial successes have been archived on the holographic dual description for the black holes, in particular for the Kerr solutions~\cite{Guica:2008mu, Dias:2009ex, Matsuo:2009sj, Bredberg:2009pv, Amsel:2009pu, Hartman:2009nz, Castro:2009jf, Cvetic:2009jn, Chen:2010ni, Castro:2010fd}, as well as the other generalizations~\cite{KerrCFT}.  The original paper on the Kerr/CFT correspondence~\cite{Guica:2008mu} explored the holographic correspondence for the extremal Kerr black holes. More precisely, it has shown that the central charge of the dual CFT can be derived from the asymptotic symmetry group of the near horizon geometry, and the temperatures can be identified via the Boltzmann factor. The major evidence is the fact that the CFT entropy computed by using the Cardy formula exactly reproduces the black hole Bekenstein-Hawking entropy. Soon after this stimulating progress, the investigation on the Kerr/CFT correspondence has been extended to the near extremal cases~\cite{Bredberg:2009pv, Hartman:2009nz} with a new support that the scattering absorption cross section of a probe field, in suitable limits, agrees with the two-point function of the dual operator. Recently, the Kerr/CFT correspondence was remarkably generalized to the generic non-extremal Kerr black holes~\cite{Castro:2010fd}. For the non-extremal black holes, the near horizon geometry does not contain an explicit AdS$_3$ structure, nevertheless, there is a local conformal invariance in the solution space of a specified probe field which merely ensures a dual CFT description. This observation indicates that even though the near-horizon geometry of a generic Kerr black hole could be distinct from the AdS or warped AdS spacetime, the local conformal symmetry on the solution space may still allow us to explore its CFT description. Both the microscopic entropy counting and the low frequency scattering amplitude in the near region support such intuitive judgement. The study of hidden conformal symmetry has been generalized to various types of black holes~\cite{HiddenSymmetry}.

For the charged black holes the holographic duality changes to have multiple faces. It has been shown in~\cite{Chen:2010ywa}, see also~\cite{Chen:2010jj}, that there are two different individual 2D CFTs holographically dual to the Kerr-Newman black holes refereing to the two possible limits: neutral Kerr and non-rotating Reissener-Nordstrom (RN) solutions. The twofold holographic descriptions, called J-picture and Q-picture, distinctly are direct extensions respectively of the Kerr/CFT correspondence and the RN/CFT correspondence~\cite{Hartman:2008pb, Garousi:2009zx, Chen:2009ht, Chen:2010bsa, Chen:2010yu}. Just like the Kerr-Newman solutions, the charged rotating black holes in the five-dimensional minimal supergravity~\cite{Chong:2005hr} can provide another interesting backgrounds to verity the validity of the holographic principle. Several profound investigations have been done, including the duality for the extremal limit, see for example~\cite{Chen:2009xja}, and the associated hidden conformal symmetry~\cite{Setare:2010sy}. However, all of the study were only focusing on the angular momentum description, namely the J-picture. It is a natural expectation that there should be a holographic Q-picture description for the black holes in the five dimensional minimal supergravity. In this article, we will explore this picture in more details.

The paper is organized as follows. In section 2, we review some basic properties of the black holes in the five dimensional minimal supergravity. In section 3, the dynamics of a probe massless charged scalar field propagating in considered black hole background is studied. We investigate the Q-picture hidden conformal symmetry by analyzing the wave equation of the probe scalar field. As expected, we confirm that the microscopic entropy evaluated by the Cardy formula exactly reproduces the black hole Bekenstein-Hawking entropy. In section 4, a further support of agreement between the absorption cross section and two point function is checked. Finally, the last section is devoted to the conclusion.

\section{Black holes in 5D minimal supergravity}
In this section we review and examine the black hole solutions in the five dimensional minimal supergravity
\begin{equation}
S_5 = \frac1{16 \pi} \left[ \int d^5x \sqrt{- g} \left( R - \frac14 F^2 \right) - \frac1{3 \sqrt3} \int F \wedge F \wedge A \right].
\end{equation}
The electric charged rotating black holes~\cite{Chong:2005hr}, in the Boyer-Lindquist coordinates $x^\mu = (t, r, \theta, \varphi_1, \varphi_2)$, have the following non-vanishing metric components
\begin{eqnarray} \label{metric}
g_{00} &=& - \left( 1 - \frac{2 m}{\rho^2} - \frac{m^2}{\rho^4} \right),
\nonumber\\
g_{03} &=& - \frac{a (2 m \rho^2 - q^2) + b q \rho^2 \sin^2\theta}{\rho^4},
\nonumber\\
g_{04} &=& - \frac{b (2 m \rho^2 - q^2) + a q \rho^2 \cos^2\theta}{\rho^4},
\nonumber\\
g_{33} &=& (r^2 + a^2) \sin^2\theta + \frac{a^2 (2 m \rho^2 - q^2) + 2 a b q \rho^2}{\rho^4} \sin^4\theta,
\nonumber\\
g_{44} &=& (r^2 + b^2) \cos^2\theta + \frac{b^2 (2 m \rho^2 - q^2) + 2 a b q \rho^2}{\rho^4} \cos^4\theta,
\nonumber\\
g_{34} &=& \frac{a b (2 m \rho^2 - q^2) + (a^2 + b^2) q \rho^2}{\rho^4} \sin^2\theta \cos^2\theta,
\nonumber\\
g_{11} &=& \frac{\rho^2}{\Delta}, \qquad g_{22} = \rho^2,
\end{eqnarray}
and the gauge potential
\begin{equation} \label{A}
A = - \frac{\sqrt3 q}{\rho^2}(dt - a \sin^2\theta d\varphi_1 - b \cos^2\theta d\varphi_2),
\end{equation}
where
\begin{equation}
\Delta = \frac{(r^2 + a^2)(r^2 + b^2) + q^2 + 2 a b q}{r^2} - 2 m.
\end{equation}
For the above spacetime geometry, the determine of metric is $\sqrt{- \det(g_{\mu\nu})} = \sqrt{-g} = r \rho^2 \sin\theta \cos\theta$, and the locations of the event horizons are given by the singularities of the metric function which are the real roots of $r^2 \Delta = 0$. These black hole solutions are characterized by four parameters $m, q, a, b$ representing respectively the mass, charge and two independent angular momenta
\begin{equation}
M = \frac{3\pi}{4} m, \quad Q = \frac{\sqrt3 \pi}{4} q, \quad J_1 = \frac{\pi}{4} (2 m a + q b), \quad J_2 = \frac{\pi}{4} (2 m b + q a).
\end{equation}
The corresponding Hawking temperature, entropy, and the angular velocities and chemical potential on the horizon are given by
\begin{eqnarray}
T_H &=& \frac{r_+^4 - (a b + q)^2}{2 \pi r_+ [ (r_+^2 + a^2)(r_+^2 + b^2) + a b q]},
\nonumber\\
S_{BH} &=& \frac{\pi^2 [ (r_+^2 + a^2)(r_+^2 + b^2) + a b q]}{2 r_+},
\nonumber\\
\Omega_1 &=& \frac{(a r_+^2 + a b^2 + b q)}{(r_+^2 + a^2)(r_+^2 + b^2) + a b q},
\nonumber\\
\Omega_2 &=& \frac{(b r_+^2 + a^2 b + a q)}{(r_+^2 + a^2)(r_+^2 + b^2) + a b q},
\nonumber\\
\mu_q &=& \frac{\sqrt3 q r_+^2}{(r_+^2 + a^2)(r_+^2 + b^2) + a b q}.
\end{eqnarray}

The central charges associated with two angular momenta, i.e. the holographic J-picture description, has been discussed in~\cite{Chen:2009xja} and the related hidden conformal symmetries was analyzed in~\cite{Setare:2010sy}. However, as pointed out in~\cite{Chen:2010ywa}, the charged rotating black holes can have another proper dual CFT description, called the Q-picture, essentially based on their charge parameters, see also~\cite{Chen:2010jj, Chen:2010yu}. In this paper, we will mainly focus on the Q-picture CFT description for the black holes in the five dimensional minimal supergravity. The central charge of Q-picture actually is independent on the angular momenta, therefore can be obtained simply from the non-rotating countparts
\begin{eqnarray}
ds_5^2 &=& - \left( 1 - \frac{2m}{r^2} + \frac{q^2}{r^4} \right) dt^2 + \left( 1 - \frac{2m}{r^2} + \frac{q^2}{r^4} \right)^{-1} dr^2 + r^2 d\Omega_3^2,
\nonumber\\
A &=& - \frac{\sqrt3 q}{r^2} dt,
\end{eqnarray}
where $d\Omega_3^2 = d\theta^2 + \sin^2\theta d\varphi_1^2 + \cos^2\theta d\varphi_2^2$. In the extremal limit $m = q$, the radius of degenerated black hole horizons is $r_0 = \sqrt{m} = \sqrt{q}$ and the related near horizon geometry can be achieved by taking the limit ($\epsilon \to 0$)
\begin{equation}
r \to r_0 + \epsilon r, \qquad t \to \frac{m}{4 \epsilon} t.
\end{equation}
As expected, the near horizon geometry has AdS$_2 \times S^3$ structure and the gauge potential is linear in the radius coordinate
\begin{eqnarray} \label{NHsol}
ds_5^2 &=& \frac{r_0^2}4 \left( r^2 dt^2 + \frac{dr^2}{r^2} \right) + r_0^2 d\Omega_3^2,
\nonumber\\
A &=& \frac{\sqrt3}2 r_0 r dt.
\end{eqnarray}
Unlike the J-picture, the central charge of Q-picture is encoded both in the metric and the gauge potential. In order to recover the AdS$_3$ structure, one should embed the near horizon solution~(\ref{NHsol}) into a six-dimensional spacetime
\begin{equation}
ds_6^2 = ds_5^2 + (dy + A)^2,
\end{equation}
which leads the following form
\begin{equation}
ds_6^2 = \Gamma \left( - r^2 dt^2 + \frac{dr^2}{r^2} + \alpha d\Omega_3^2 \right) + \gamma (dy^2 + k r dt)^2,
\end{equation}
with
\begin{equation}
\Gamma = \frac{r_0^2}{4}, \qquad \alpha = 4, \qquad \gamma = 1, \qquad k = \frac{\sqrt3}{2} r_0.
\end{equation}
The left-sector central charge and temperature are given by
\begin{equation}
c_L = \frac{3 k}{2 \pi} \int d\Omega_3 \sqrt{(\Gamma \alpha)^3} = 3 \pi r_0^3 k, \qquad T_L =\frac{1}{2 \pi k},
\end{equation}
assuming the periodicity of coordinate $y$ is $2\pi$. The Cardy formula for CFT entropy exactly reproduces the black hole entropy
\begin{equation}
S_{CFT} = \frac{\pi^2}{3} c_L T_L = \frac12 \pi^2 r_0^3 = S_{BH}.
\end{equation}
Nevertheless, the Q-picture central charge is ambiguous up to the radius, $\ell$, of the extra circle, namely $y \sim y + 2 \pi \ell$ periodicity~\cite{Chen:2009ht, Chen:2010bsa}. Moreover, both the left-sector and right-sector central charges should be identical. Therefore, the general formula of the central charges is
\begin{equation} \label{cc}
c_L = c_R = \frac{3 \sqrt3 \pi q^2}{2 \ell}.
\end{equation}
There are two natural choices for the value of $\ell$: one is $\ell = 1$ (Planck length) and the other is $\ell = r_0$ (about the size of the AdS$_3$). From the brane construction point of view, the later choice corresponds to the configuration of long strings winding on the large extra circle~\cite{Guica:2010ej}.

\section{Q-picture hidden conformal symmetry}
In this section, we consider the Klein-Gordon (KG) equation for a probe complex scalar field propagating in the considered black hole background. For a massless complex scalar field carrying the charge $e$, the KG equation
\begin{equation}
(\nabla_\mu - i e A_\mu) (\nabla^\mu - i e A^\mu) \Phi = 0,
\end{equation}
can be simplified by assuming the following form of the scalar field
\begin{eqnarray}
\Phi = \exp(-i \omega t + i m_1 \varphi_1 + i m_2 \varphi_2) S(\theta) R(r).
\end{eqnarray}
The neutral scalar field ($e = 0$) is able to reveal, from the radial equation, the hidden conformal symmetry (in J-picture) as shown in~\cite{Setare:2010sy}. For the Q-picture description we should consider the radial equation with conditions $m_1 = m_2 = 0$~\cite{Chen:2010ywa, Chen:2010yu}, so the radial equation can be expressed as
\begin{equation}
\frac{1}{r} \partial_r ( r \Delta \partial_r R ) + \left[ \frac{\left( [(r^2 + a^2)(r^2 + b^2) + a b q] \omega - \sqrt{3} e q r^2 \right)^2}{r^4 \Delta} - \frac{a^2 b^2 \omega^2}{r^2} - \lambda \right] R = 0.
\end{equation}
Defining a new variable $u = r^2$ and $u_+= r_+^2, \; u_- = r_-^2$, then
\begin{equation}
\tilde \Delta \equiv r^2 \Delta = (u - u_+) (u - u_-),
\end{equation}
and the above differential equation becomes
\begin{equation}
4 \partial_u ( \tilde\Delta \partial_u R ) + \left[ \frac{\left( [(u + a^2)(u + b^2) + a b q] \omega - \sqrt{3} e q u \right)^2}{u (u - u_+) (u - u_-)} - \frac{a^2 b^2 \omega^2}{u} - \lambda \right] R = 0.
\end{equation}
In the limits for the scalar field with low frequency $\omega^2 r_+ \ll 1$ (consequently $\omega^2 m^2 \ll 1, \omega a \ll 1, \omega b \ll 1$) and small charge $e q \ll 1$, the radial equation in the near region $r \omega \ll 1$ could be simplified as\footnote{The separation constant $\lambda$ reduces to $l (l + 2)$ corresponding to the spherical harmonics of the $S^3$.}
\begin{eqnarray} \label{EqRlim}
\left[ \partial_u (\tilde\Delta \partial_u) + \frac{(\beta_+ \omega - \sqrt{3} e q r_+)^2}{4 (u - u_+) (u_+ - u_-)} - \frac{(\beta_- \omega - \sqrt{3} e q r_-)^2}{4 (u - u_-) (u_+ - u_-)} \right] R = \frac{l (l + 2)}4 R,
\end{eqnarray}
where
\begin{equation} \label{beta}
\beta_\pm = \frac{(r_\pm^2 + a^2) (r_\pm ^2 + b^2) + a b q}{r_\pm}.
\end{equation}
Following the idea proposed in~\cite{Castro:2010fd} we are going to show that the equation~(\ref{EqRlim}) actually can be reproduced by the Casimir operator of the AdS$_3$ space
\begin{equation} \label{dsAdS3}
ds_3^2 = \frac{L^2}{y^2} (dy^2 + dw^+ dw^-).
\end{equation}
Here, the AdS$_3$ radius $L$ is not essential in our discussion. There are two sets of symmetry generators
\begin{equation}
H_1 = i \partial_+, \quad H_0 = i (w^+ \partial_+ + \frac{1}{2} y \partial_y), \quad H_{-1} = i ( (w^+)^2 \partial_+ + w^+ y \partial_y - y^2 \partial_-),
\end{equation}
and
\begin{equation}
\bar{H}_1 = i \partial_-, \quad \bar{H}_0 = i (w^- \partial_- + \frac{1}{2} y \partial_y), \quad \bar{H}_{-1} = i ( (w^-)^2 \partial_- + w^- y \partial_y - y^2 \partial_+),
\end{equation}
each of them satisfies the $SL(2, R)$ algebra
\begin{equation}
[H_0, H_{\pm 1}] = \mp i H_{\pm 1}, \quad [H_{-1}, H_1] = -2 i H_0,
\end{equation}
and
\begin{equation}
[\bar{H}_0, \bar{H}_{\pm 1}] = \mp i \bar{H}_{\pm 1}, \quad [\bar{H}_{-1}, \bar{H}_1] = -2 i \bar{H}_0.
\end{equation}
Coordinately, the associated quadratic Casimir operator is
\begin{eqnarray}
{\cal H}^2 = \bar{\cal H}^2 = - H_0^2 + \frac{1}{2} (H_1 H_{-1} + H_{-1} H_1) = \frac{1}{4} (y^2 \partial_y^2 - y\partial_y) + y^2 \partial_+ \partial_-.
\end{eqnarray}
By introducing the following transformations from the conformal space to black hole coordinates
\begin{eqnarray} \label{C2BH}
w^+ &=& \sqrt{\frac{u - u_+}{u - u_-}} \exp(2 \pi T_R \chi + 2 n_R t),
\nonumber\\
w^- &=& \sqrt{\frac{u - u_+}{u - u_-}} \exp(2 \pi T_L \chi + 2 n_L t),
\\
y &=& \sqrt{\frac{u_+ - u_-}{u - u_-}} \exp(\pi (T_R + T_L) \chi + (n_R + n_L) t),
\nonumber
\end{eqnarray}
the Casimir operator is transformed in terms of $(u, t, \chi)$ coordinates as
\begin{eqnarray}\label{Casimir}
{\cal H}^2 &=& \partial_u (\tilde\Delta \partial_u) - \frac{u_+ - u_-}{u - u_+} \left( \frac{T_L + T_R}{4 {\cal A}} \partial_t - \frac{n_L + n_R}{4 \pi {\cal A}} \partial_\chi \right)^2
\nonumber\\
&& + \frac{u_+ - u_-}{u - u_-} \left( \frac{T_L - T_R}{{4 \cal A}} \partial_t - \frac{n_L - n_R}{4 \pi {\cal A}} \partial_\chi \right)^2,
\end{eqnarray}
where ${\cal A} = T_R n_L - T_L n_R$. Furthermore, from the black hole side, the radial equation~(\ref{EqRlim}) can be reexpressed as
\begin{eqnarray} \label{EqR}
\left[ \partial_u (\tilde\Delta \partial_u) - \frac{(\beta_+ \partial_t - (\sqrt3 q r_+/\ell) \partial_\chi)^2}{4 (u - u_+) (u_+ - u_-)} + \frac{(\beta_- \partial_t - (\sqrt3 q r_-/\ell) \partial_\chi)^2}{4 (u - u_-) (u_+ - u_-)} \right] R = \frac{l (l + 2)}4 R,
\end{eqnarray}
after introducing an operator $\partial_\chi$ acting on the $U(1)$ symmetry internal space of the complex scalar field. The eigenvalue of the new operator is the scalar field charge~\cite{Chen:2010ywa, Chen:2010yu}, namely $\partial_\chi \Phi = i \ell e \Phi$, up to an undetermined parameter $\ell$ correlated with the ambiguity in the central charge. Therefore, the radial equation can be realized as the Casimir operator~(\ref{Casimir}) acting on $\Phi$ with the following identifications, including the CFT temperatures
\begin{eqnarray}\label{TLR}
T_L = \frac{\ell(\beta_+ + \beta_-)}{2 \sqrt3 \pi q^2}, &\quad& T_R = \frac{\ell(\beta_+ - \beta_-)}{2 \sqrt3 \pi q^2},
\\
n_L = \frac{r_+ + r_-}{2 q}, &\quad& n_R = \frac{r_+ - r_-}{2 q}.
\end{eqnarray}
As the first evidence, one can easily verity the agreement of the microscopic and macroscopic entropies
\begin{equation}
S_{CFT} = \frac{\pi^2}3 (c_L T_L + c_R T_R) = \frac{\pi^2}{2} \beta_+ = S_{BH}.
\end{equation}

\section{Absorption cross section}
For a further support to the holographic Q-picture, we will show that the absorption cross section for the probe scalar field (with assumptions $m_1 = m_2 = 0$) scattered in the near region of the black hole matches with the two point function of the dual operator in the CFT with identified, left and right, central charges and temperatures. The absorption cross section can be written as~\cite{Chen:2010ywa, Cvetic:1997uw}
\begin{equation}
P_\mathrm{abc} \sim \sinh(2 \pi \gamma_Q) \left| \Gamma(a_Q) \right|^2 \left| \Gamma(b_Q) \right|^2
\end{equation}
where the three coefficients can be straightforwardly read out from the equation~(\ref{EqRlim})
\begin{eqnarray}
\gamma_Q &=& \frac{\beta_+ \omega - \sqrt3 e q r_+}{2 (r_+^2 - r_-^2)},
\nonumber\\
a_Q &=& 1 + \frac{l}2 - i \frac{(\beta_+ + \beta_-) \omega - \sqrt3 e q (r_+ + r_-)}{2 (r_+^2 - r_-^2)},
\nonumber\\
b_Q &=& 1 + \frac{l}2 - i \frac{(\beta_+ - \beta_-) \omega - \sqrt3 e q (r_+ - r_-)}{2 (r_+^2 - r_-^2)},
\end{eqnarray}
leading to the relation $a_Q + b_Q = 2 + l - 2 i \gamma_Q$. In order to explicitly check that the $P_\mathrm{abs}$ really matches with the microscopic greybody factor of the dual CFT, one needs to identify the related parameters of the dual operator. Firstly, the conformal weights of the dual operator is
\begin{equation}
(h_L, h_R) = \left( 1 + \frac{l}{2}, 1 + \frac{l}{2} \right).
\end{equation}
Moreover, from the first law of black hole thermodynamics
\begin{equation}
T_H \delta S_{BH} = \delta m - \Omega_1 \delta J_1 - \Omega_2 \delta J_2 - \mu_q \delta q,
\end{equation}
one can identify the conjugate charges as
\begin{equation}
\delta S_{BH} = \delta S_{CFT} = \frac{\delta E_L}{T_L} + \frac{\delta E_R}{T_R}.
\end{equation}
In the Q-picture description, one should assume $\delta m = \omega, \delta q = e$ and $\delta J_1 = \delta J_2 = 0$, and the solution of the conjugate charges is
\begin{equation}
\delta E_L = \tilde\omega_L = \omega_L - q_L \mu_L, \qquad \delta E_R = \tilde\omega_R = \omega_R - q_R \mu_R,
\end{equation}
where
\begin{eqnarray}
\omega_L = \frac{\ell (\beta_+^2 - \beta_-^2)}{2 \sqrt3 q^2 (r_+^2 - r_-^2)} \omega, \qquad \mu_L = \frac{\ell (\beta_+ + \beta_-)}{2 q (r_+ + r_-)}, \qquad q_L = e,
\nonumber\\
\omega_R = \frac{\ell (\beta_+^2 - \beta_-^2)}{2 \sqrt3 q^2 (r_+^2 - r_-^2)} \omega, \qquad \mu_R = \frac{\ell (\beta_+ - \beta_-)}{2 q (r_+ - r_-)}, \qquad q_R = e.
\end{eqnarray}
Finally, the absorption cross section can be expressed as
\begin{equation}
P_\mathrm{abs} \sim T_L^{2 h_L - 1} T_R^{2 h_R - 1} \sinh\left( \frac{\tilde\omega_L}{2 T_L} + \frac{\tilde\omega_R}{2 T_R} \right) \left| \Gamma\left( h_L + i \frac{\tilde\omega_L}{2 \pi T_L} \right) \right|^2 \left| \Gamma\left( h_R + i \frac{\tilde\omega_R}{2 \pi T_R} \right) \right|^2,
\end{equation}
in agreement with the two point function of the dual operator.

\section{Conclusion}
The ``microscopic hair conjecture'' proposed in~\cite{Chen:2010ywa} claims that for each macroscopic hair parameter, in additional
to the mass, of a black hole there should exist an associated holographic CFT$_2$ description. For the charged rotating black holes in the five dimensional minimal supergravity, the J-picture descriptions associated with two angular momenta has been studied previously  in~\cite{Chen:2009xja} for the extremal case and in~\cite{Setare:2010sy} for the hidden conformal symmetry. In this paper, we have explored the supplementary holographic description, the Q-picture, based on the electric charge of the black hole. The central charge of the Q-picture CFT actually is independent on the angular momenta, so it can be computed simply from the non-rotating countparts of black hole. Unlike the J-picture, the central charge of Q-picture CFT is encoded both in the metric and the gauge potential. The charge contribution can not be obtained directly from the central extension of the asymptotic symmetry group~\cite{Compere:2009dp}. In generic, the near horizon geometry of an extremal non-rotating charged black hole has only an AdS$_2$ structure and the $U(1)$ fiber of the fundamental AdS$_3$ is held by the gauge potential which can been revealed by a Kaluza-Klein uplifting.

Specifically, we consider the wave equation of a massless charged scalar field in the background of black holes in the five dimensional minimal supergravity. It turns out that under certain low frequency and low charge limits, the radial part of the ``near region'' KG equation is equivalent to a Casimir operator of $SL(2,R)_L \times SL(2,R)_R$ group. The CFT temperatures then can be identified. The macroscopic entropy and the absorption cross section of the probe scalar field match precisely to the microscopic CFT entropy and the corresponding two point function. All our results provide evidences for the validity of the holographic Q-picture description of the black holes in the five dimensional minimal supergravity.

\section*{Acknowledgements}
This work by CMC was supported by the National Science Council of the R.O.C. under the grant NSC 99-2112-M-008-005-MY3 and in part by the National Center of Theoretical Sciences (NCTS).



\begin{thebibliography}{99}


\bibitem{'tHooft:1993gx}
  G.~'t Hooft,
  ``Dimensional reduction in quantum gravity,''
  arXiv:gr-qc/9310026.

\bibitem{Susskind:1994vu}
  L.~Susskind,
  ``The world as a hologram,''
  J.\ Math.\ Phys.\  {\bf 36}, 6377 (1995)
  [arXiv:hep-th/9409089].

\bibitem{Maldacena:1997re}
  J.~M.~Maldacena,
  ``The large N limit of superconformal field theories and supergravity,''
  Adv.\ Theor.\ Math.\ Phys.\  {\bf 2}, 231 (1998)
  [Int.\ J.\ Theor.\ Phys.\  {\bf 38}, 1113 (1999)]
  [arXiv:hep-th/9711200].


\bibitem{Guica:2008mu}
  M.~Guica, T.~Hartman, W.~Song and A.~Strominger,
  ``The Kerr/CFT Correspondence,''
  Phys.\ Rev.\  D {\bf 80}, 124008 (2009)
  [arXiv:0809.4266 [hep-th]].

\bibitem{Dias:2009ex}
  O.~J.~C.~Dias, H.~S.~Reall and J.~E.~Santos,
  ``Kerr-CFT and gravitational perturbations,''
  JHEP {\bf 0908}, 101 (2009)
  [arXiv:0906.2380 [hep-th]].

\bibitem{Matsuo:2009sj}
  Y.~Matsuo, T.~Tsukioka and C.~M.~Yoo,
  ``Another Realization of Kerr/CFT Correspondence,''
  Nucl.\ Phys.\  B {\bf 825}, 231 (2010)
  [arXiv:0907.0303 [hep-th]].

\bibitem{Bredberg:2009pv}
  I.~Bredberg, T.~Hartman, W.~Song and A.~Strominger,
  ``Black Hole Superradiance From Kerr/CFT,''
  JHEP {\bf 1004}, 019 (2010)
  [arXiv:0907.3477 [hep-th]].

\bibitem{Amsel:2009pu}
  A.~J.~Amsel, D.~Marolf and M.~M.~Roberts,
  ``On the Stress Tensor of Kerr/CFT,''
  JHEP {\bf 0910}, 021 (2009)
  [arXiv:0907.5023 [hep-th]].

\bibitem{Hartman:2009nz}
  T.~Hartman, W.~Song and A.~Strominger,
  ``Holographic Derivation of Kerr-Newman Scattering Amplitudes for General Charge and Spin,''
  JHEP {\bf 1003}, 118 (2010)
  [arXiv:0908.3909 [hep-th]].

\bibitem{Castro:2009jf}
  A.~Castro and F.~Larsen,
  ``Near Extremal Kerr Entropy from AdS$_2$ Quantum Gravity,''
  JHEP {\bf 0912}, 037 (2009)
  [arXiv:0908.1121 [hep-th]].

\bibitem{Cvetic:2009jn}
  M.~Cvetic and F.~Larsen,
  ``Greybody Factors and Charges in Kerr/CFT,''
  JHEP {\bf 0909}, 088 (2009)
  [arXiv:0908.1136 [hep-th]].

\bibitem{Chen:2010ni}
  B.~Chen and C.~S.~Chu,
  ``Real-time correlators in Kerr/CFT correspondence,''
  JHEP {\bf 1005}, 004 (2010)
  [arXiv:1001.3208 [hep-th]].

\bibitem{Castro:2010fd}
  A.~Castro, A.~Maloney and A.~Strominger,
  ``Hidden Conformal Symmetry of the Kerr Black Hole,''
  Phys.\ Rev.\  D {\bf 82}, 024008 (2010)
  [arXiv:1004.0996 [hep-th]].



\bibitem{KerrCFT}
  T.~Azeyanagi, N.~Ogawa and S.~Terashima,
  ``Holographic Duals of Kaluza-Klein Black Holes,''
  JHEP {\bf 0904}, 061 (2009)
  [arXiv:0811.4177 [hep-th]].
\\
  D.~D.~K.~Chow, M.~Cvetic, H.~Lu and C.~N.~Pope,
  ``Extremal Black Hole/CFT Correspondence in (Gauged) Supergravities,''
  Phys.\ Rev.\  D {\bf 79}, 084018 (2009)
  [arXiv:0812.2918 [hep-th]]..
\\
  H.~Isono, T.~S.~Tai and W.~Y.~Wen,
  ``Kerr/CFT correspondence and five-dimensional BMPV black holes,''
  Int.\ J.\ Mod.\ Phys.\  A {\bf 24}, 5659 (2009)
  [arXiv:0812.4440 [hep-th]].
\\
  T.~Azeyanagi, N.~Ogawa and S.~Terashima,
  ``The Kerr/CFT Correspondence and String Theory,''
  Phys.\ Rev.\  D {\bf 79}, 106009 (2009)
  [arXiv:0812.4883 [hep-th]].
\\
  J.~J.~Peng and S.~Q.~Wu,
  ``Extremal Kerr black hole/CFT correspondence in the five dimensional G\'odel universe,''
  Phys.\ Lett.\  B {\bf 673}, 216 (2009)
  [arXiv:0901.0311 [hep-th]].
\\
  F.~Loran and H.~Soltanpanahi,
  ``5D Extremal Rotating Black Holes and CFT duals,''
  Class.\ Quant.\ Grav.\  {\bf 26}, 155019 (2009)
  [arXiv:0901.1595 [hep-th]].
\\
  A.~M.~Ghezelbash,
  ``Kerr/CFT Correspondence in Low Energy Limit of Heterotic String Theory,''
  JHEP {\bf 0908}, 045 (2009)
  [arXiv:0901.1670 [hep-th]].
\\
  H.~Lu, J.~w.~Mei, C.~N.~Pope and J.~F.~Vazquez-Poritz,
  ``Extremal Static AdS Black Hole/CFT Correspondence in Gauged Supergravities,''
  Phys.\ Lett.\  B {\bf 673}, 77 (2009)
  [arXiv:0901.1677 [hep-th]].
\\
  K.~Hotta,
  ``Holographic RG flow dual to attractor flow in extremal black holes,''
  Phys.\ Rev.\  D {\bf 79}, 104018 (2009)
  [arXiv:0902.3529 [hep-th]].
\\
  D.~Astefanesei and Y.~K.~Srivastava,
  ``CFT Duals for Attractor Horizons,''
  Nucl.\ Phys.\  B {\bf 822}, 283 (2009)
  [arXiv:0902.4033 [hep-th]].
\\
  A.~M.~Ghezelbash,
  ``Kerr-Bolt Spacetimes and Kerr/CFT Correspondence,''
  arXiv:0902.4662 [hep-th].
\\
  C.~Krishnan and S.~Kuperstein,
  ``A Comment on Kerr-CFT and Wald Entropy,''
  arXiv:0903.2169 [hep-th].
\\
  T.~Azeyanagi, G.~Compere, N.~Ogawa, Y.~Tachikawa and S.~Terashima,
  ``Higher-Derivative Corrections to the Asymptotic Virasoro Symmetry of 4d Extremal Black Holes,''
  Prog.\ Theor.\ Phys.\  {\bf 122}, 355 (2009)
  [arXiv:0903.4176 [hep-th]].
\\
  X.~N.~Wu and Y.~Tian,
  ``Extremal Isolated Horizon/CFT Correspondence,''
  Phys.\ Rev.\  D {\bf 80}, 024014 (2009)
  [arXiv:0904.1554 [hep-th]].
\\
  J.~Rasmussen,
  ``Isometry-preserving boundary conditions in the Kerr/CFT correspondence,''
  Int.\ J.\ Mod.\ Phys.\  A {\bf 25}, 1597 (2010)
  [arXiv:0908.0184 [hep-th]].


\bibitem{HiddenSymmetry}
  C.~Krishnan,
  ``Hidden Conformal Symmetries of Five-Dimensional Black Holes,''
  JHEP {\bf 1007}, 039 (2010)
  [arXiv:1004.3537 [hep-th]].
\\
  Y.~Q.~Wang and Y.~X.~Liu,
  ``Hidden Conformal Symmetry of the Kerr-Newman Black Hole,''
  JHEP {\bf 1008}, 087 (2010)
  [arXiv:1004.4661 [hep-th]].
\\
B.~Chen and J.~Long,
  ``Real-time Correlators and Hidden Conformal Symmetry in Kerr/CFT Correspondence,''
  JHEP {\bf 1006}, 018 (2010)
  [arXiv:1004.5039 [hep-th]].
\\
  R.~Li, M.~F.~Li and J.~R.~Ren,
  ``Entropy of Kaluza-Klein Black Hole from Kerr/CFT Correspondence,''
  Phys.\ Lett.\  B {\bf 691}, 249 (2010)
  [arXiv:1004.5335 [hep-th]].
\\
  D.~Chen, P.~Wang and H.~Wu,
  ``Hidden conformal symmetry of rotating charged black holes,''
  arXiv:1005.1404 [gr-qc].
\\
  M.~Becker, S.~Cremonini and W.~Schulgin,
  ``Correlation Functions and Hidden Conformal Symmetry of Kerr Black Holes,''
  JHEP {\bf 1009}, 022 (2010)
  [arXiv:1005.3571 [hep-th]].
\\
  B.~Chen and J.~Long,
  ``On Holographic description of the Kerr-Newman-AdS-dS black holes,''
  JHEP {\bf 1008}, 065 (2010)
  [arXiv:1006.0157 [hep-th]].
\\
  R.~Fareghbal,
  ``Hidden Conformal Symmetry of Warped AdS$_3$ Black Holes,''
  Phys.\ Lett.\  B {\bf 694}, 138 (2010)
  [arXiv:1006.4034 [hep-th]].
\\
  H.~Wang, D.~Chen, B.~Mu and H.~Wu,
  ``Hidden conformal symmetry of extreme and non-extreme Einstein-Maxwell-Dilaton-Axion black holes,''
  JHEP {\bf 1011}, 002 (2010)
  [arXiv:1006.0439 [gr-qc]].
\\
  R.~Li, M.~F.~Li and J.~R.~Ren,
  ``Hidden Conformal Symmetry of Self-Dual Warped AdS$_3$ Black Holes in Topological Massive Gravity,''
  arXiv:1007.1357 [hep-th].
\\
  Y.~Matsuo, T.~Tsukioka and C.~M.~Yoo,
  ``Notes on the Hidden Conformal Symmetry in the Near Horizon Geometry of the Kerr Black Hole,''
  arXiv:1007.3634 [hep-th].
\\
  K.~N.~Shao and Z.~Zhang,
  ``Hidden Conformal Symmetry of Rotating Black Hole with four Charges,''
  arXiv:1008.0585 [hep-th].
\\
  A.~M.~Ghezelbash, V.~Kamali and M.~R.~Setare,
  ``Hidden Conformal Symmetry of Kerr-Bolt Spacetimes,''
  arXiv:1008.2189 [hep-th].
\\
  B.~Chen, A.~M.~Ghezelbash, V.~Kamali and M.~R.~Setare,
  ``Holographic description of Kerr-Bolt-AdS-dS Spacetimes,''
  arXiv:1009.1497 [hep-th].
\\
  R.~li and J.~R.~Ren,
  ``Holographic Dual of Linear Dilaton Black Hole in Einstein-Maxwell-Dilaton-Axion Gravity,''
  JHEP {\bf 1009}, 039 (2010)
  [arXiv:1009.3139 [hep-th]].
\\
  M.~R.~Setare and V.~Kamali,
  ``Hidden Conformal Symmetry of Extremal Kerr-Bolt Spacetimes,''
  JHEP {\bf 1010}, 074 (2010)
  [arXiv:1011.0809 [hep-th]].



\bibitem{Chen:2010ywa}
  C.~M.~Chen, Y.~M.~Huang, J.~R.~Sun, M.~F.~Wu and S.~J.~Zou,
  ``Twofold Hidden Conformal Symmetries of the Kerr-Newman Black Hole,''
  Phys.\ Rev.\  D {\bf 82}, 066004 (2010)
  [arXiv:1006.4097 [hep-th]].

\bibitem{Chen:2010jj}
  B.~Chen, C.~M.~Chen and B.~Ning,
  ``Holographic Q-picture of Kerr-Newman-AdS-dS Black Hole,''
  arXiv:1010.1379 [hep-th].



\bibitem{Hartman:2008pb}
  T.~Hartman, K.~Murata, T.~Nishioka and A.~Strominger,
  ``CFT Duals for Extreme Black Holes,''
  JHEP {\bf 0904}, 019 (2009)
  [arXiv:0811.4393 [hep-th]].

\bibitem{Garousi:2009zx}
  M.~R.~Garousi and A.~Ghodsi,
  ``The RN/CFT Correspondence,''
  Phys.\ Lett.\  B {\bf 687}, 79 (2010)
  [arXiv:0902.4387 [hep-th]].

\bibitem{Chen:2009ht}
  C.~M.~Chen, J.~R.~Sun and S.~J.~Zou,
  ``The RN/CFT Correspondence Revisited,''
  JHEP {\bf 1001}, 057 (2010)
  [arXiv:0910.2076 [hep-th]].

\bibitem{Chen:2010bsa}
  C.~M.~Chen, Y.~M.~Huang and S.~J.~Zou,
  ``Holographic Duals of Near-extremal Reissner-Nordstrom Black Holes,''
  JHEP {\bf 1003}, 123 (2010)
  [arXiv:1001.2833 [hep-th]].

\bibitem{Chen:2010yu}
  C.~M.~Chen, Y.~M.~Huang, J.~R.~Sun, M.~F.~Wu and S.~J.~Zou,
  ``On Holographic Dual of the Dyonic Reissner-Nordstr\'om Black Hole,''
  Phys.\ Rev.\  D {\bf 82}, 066003 (2010)
  [arXiv:1006.4092 [hep-th]].



\bibitem{Chong:2005hr}
  Z.~W.~Chong, M.~Cvetic, H.~Lu and C.~N.~Pope,
  ``General non-extremal rotating black holes in minimal five-dimensional gauged supergravity,''
  Phys.\ Rev.\ Lett.\  {\bf 95}, 161301 (2005)
  [arXiv:hep-th/0506029].

\bibitem{Chen:2009xja}
  C.~M.~Chen and J.~E.~Wang,
  ``Holographic Duals of Black Holes in Five-dimensional Minimal Supergravity,''
  Class.\ Quant.\ Grav.\  {\bf 27}, 075004 (2010)
  [arXiv:0901.0538 [hep-th]].

\bibitem{Setare:2010sy}
  M.~R.~Setare and V.~Kamali,
  ``Hidden Conformal Symmetry of Rotating Black Holes in Minimal Five-Dimensional Gauged Supergravity,''
  Phys.\ Rev.\  D {\bf 82}, 086005 (2010)
  [arXiv:1008.1123 [hep-th]].


\bibitem{Guica:2010ej}
  M.~Guica and A.~Strominger,
  ``Microscopic Realization of the Kerr/CFT Correspondence,''
  arXiv:1009.5039 [hep-th].

\bibitem{Cvetic:1997uw}
  M.~Cvetic and F.~Larsen,
  ``General rotating black holes in string theory: Greybody factors and event horizons,''
  Phys.\ Rev.\  D {\bf 56}, 4994 (1997)
  [arXiv:hep-th/9705192].

\bibitem{Compere:2009dp}
  G.~Compere, K.~Murata and T.~Nishioka,
  ``Central Charges in Extreme Black Hole/CFT Correspondence,''
  JHEP {\bf 0905}, 077 (2009)
  [arXiv:0902.1001 [hep-th]].

\end{thebibliography}
\end{document}